\documentclass[
reprint, amsmath,amssymb,aps,prl,superscriptaddress
]{revtex4-1}

\usepackage{graphicx}
\usepackage{amsmath} 
\usepackage{amsfonts}
\usepackage{bbm}
\usepackage{braket}
\usepackage{caption}
\usepackage{graphicx}
\usepackage{float}
\usepackage{subcaption}
\usepackage{url}
\usepackage{qcircuit}
\usepackage{amssymb}
\usepackage{qcircuit}
\usepackage{caption}
\usepackage{siunitx}
\usepackage{braket}
\usepackage{multirow}
\usepackage{color}

\begin{document}

\title{High-fidelity Two-qubit Gates Using a MEMS-based Beam Steering System for Individual Qubit Addressing}

\author{Ye Wang}
\email{ye.wang2@duke.edu}
\affiliation{Department of Electrical and Computer Engineering, Duke University, Durham, NC 27708, USA}
\author{Stephen Crain}
\affiliation{Department of Electrical and Computer Engineering, Duke University, Durham, NC 27708, USA}
\author{Chao Fang}
\affiliation{Department of Electrical and Computer Engineering, Duke University, Durham, NC 27708, USA}
\author{Bichen Zhang}
\affiliation{Department of Electrical and Computer Engineering, Duke University, Durham, NC 27708, USA}
\author{Shilin Huang}
\affiliation{Department of Electrical and Computer Engineering, Duke University, Durham, NC 27708, USA}
\author{Qiyao Liang}
\affiliation{Department of Physics, Duke University, Durham, NC 27708, USA}
\author{Pak Hong Leung}
\affiliation{Department of Physics, Duke University, Durham, NC 27708, USA}
\author{Kenneth R. Brown}
\email{ken.r.brown@duke.edu}
\affiliation{Department of Electrical and Computer Engineering, Duke University, Durham, NC 27708, USA}
\affiliation{Department of Physics, Duke University, Durham, NC 27708, USA}
\affiliation{Department of Chemistry, Duke University, Durham, NC 27708, USA}

\author{Jungsang Kim}
\email{jungsang.kim@duke.edu}
\affiliation{Department of Electrical and Computer Engineering, Duke University, Durham, NC 27708, USA}
\affiliation{IonQ, Inc., College Park, MD 20740, USA}

\begin{abstract} 
In a large scale trapped atomic ion quantum computer, high-fidelity two-qubit gates need to be extended over all qubits with individual control. We realize and characterize high-fidelity two-qubit gates in a system with up to 4 ions using radial modes. The ions are individually addressed by two tightly focused beams steered using micro-electromechanical system (MEMS) mirrors. We deduce a gate fidelity of $99.49(7) \%$ in a two-ion chain and $99.30(6) \%$ in a four-ion chain by applying a sequence of up to 21 two-qubit gates and measuring the final state fidelity. We characterize the residual errors and discuss methods to further improve the gate fidelity towards values that are compatible with fault-tolerant quantum computation.

\end{abstract}

\maketitle

Trapped atomic ions are one of the leading qubit platforms for realizing a quantum computer due to long coherence times \cite{wang2017single} and high-fidelity initialization, detection, and qubit gate operation \cite{crain2019high,ballance2016high,gaebler2016high,harty2014high,pino2020demonstration}. The M\o lmer-S\o rensen (MS) gate \cite{sorensen1999quantum,sorensen2000entanglement} is a widely used two-qubit gate with demonstrated fidelities above 99.9$\%$ in two-ion systems utilizing axial modes~\cite{ballance2016high,gaebler2016high}. For practical applications such as digital quantum simulation \cite{lanyon2011universal} and fault-tolerant quantum computation \cite{shor1995scheme,kitaev2003fault,aharonov1999fault}, the high-fidelity two-qubit gate needs to be extended to all qubits in the system with the ability to address individual qubits. 

Individual addressing of atomic qubits in an array to realize qubit control has been accomplished by multi-channel acousto-optic modulators \cite{debnath2016demonstration, wright2019benchmarking}, steering beams using acousto/electro-optic modulators \cite{schmidt2003nature, yavuz2006fast}, and micro-electromechanical system (MEMS) tilting mirrors \cite{KnoernschildAPL2010,crain2014individual}.
For high-fidelity quantum logic gate operations in a larger array of qubits, one must consider loss of optical phase coherence 
between individual addressing beams and the crosstalk from an addressing beam to neighboring qubits that can impact the gate fidelity.
Negligible crosstalk has been demonstrated using a MEMS-based individual addressing system \cite{crain2014individual}, and gate schemes that are not sensitive to optical phase drift between the addressing beams have been developed~\cite{LeeJOB2005,InlekPRA2014,TanNature2015} to overcome the fluctuation in optical beam paths among different beams. 

Modulated pulse techniques are used to disentangle the internal qubit states from all collective motional modes and increase the robustness against frequency drifts. Amplitude-modulated (AM) gates \cite{zhu2006arbitrary, roos2008ion,debnath2016demonstration, wright2019benchmarking}, phase-modulated (PM) gates \cite{green2015phase, milne2020phase, lu2019global}, multitone MS gates \cite{shapira2018robust,blumel2019power}, and frequency-modulated (FM) gates \cite{leung2018robust,leung2018arbitrary,landsman2019two} have been developed and demonstrated. The fidelity of the AM, PM and FM gates demonstrated in a chain of five (or more) ions is around $97 \% \sim 98.5 \%$, when radial motional modes are used for the gate \cite{debnath2016demonstration, wright2019benchmarking, lu2019global, leung2018robust,landsman2019two}. 
Here, we develop the discrete FM gate, which is compatible with simple direct digital synthesizers (DDS).

With an optimized automatic calibration pipeline for the trapped ion system, we demonstrate high-fidelity two-qubit gates in a system with up to 4 ions using MEMS-based individual qubit addressing system. The two-qubit gate fidelity is $99.49(7) \%$ in a two-ion chain and $99.30(6)\%$ in a four-ion chain. 
The residual errors are analyzed and point to future directions for designing a high-fidelity two-qubit gate in longer ion chains.

The qubit is encoded in the hyperfine levels of the $^2$S$_{1/2}$ ground state manifold in a $^{171}$Yb$^+$ ion as $\ket{0} \equiv \ket{F=0;m_F = 0}$ and $\ket{1} \equiv \ket{F=1;m_F = 0}$ with a qubit frequency splitting of 12.642821 GHz \cite{olmschenk2007manipulation}, as shown in Fig. 1(a). The qubit coherence time is measured to be more than 1 second using microwave single-qubit operations with a spin echo pulse. The qubit coherence time can be extended to more than 10 minutes using well-designed dynamical decoupling pulses \cite{wang2017single}. The linear chain of $^{171}$Yb$^+$ ions are confined in a microfabricated linear radio-frequency (RF) Paul trap \cite{maunz2016high} inside an ultra-high vacuum chamber. The two radial trap frequencies are $\nu_1 = 3.1$ MHz and $\nu_2 = 2.7$ MHz. The radial principal axes are rotated about $45^{\circ}$ to the surface of the trap. The axial trap frequency is 600 kHz and 150 kHz for 2-ion and 4-ion chains with \SI{5}{\micro m} ion spacing, respectively. 

\begin{figure}[!htb]
\center{\includegraphics[width=0.48 \textwidth]{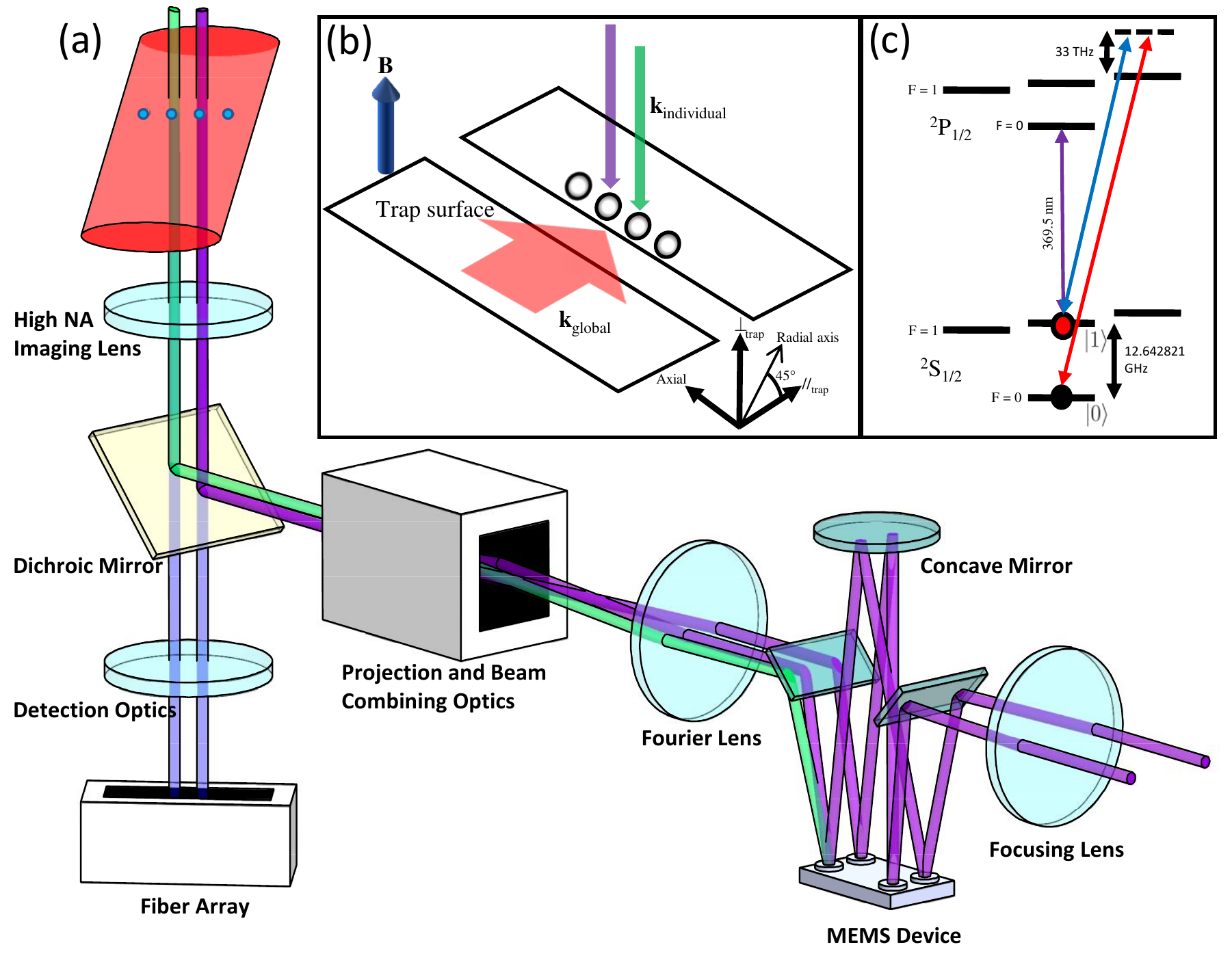}}
\caption{\label{MEMS setup} \textbf{(a), (b)} Schematic representation of the Raman beam optical setup. The two individual addressing beams (purple) are steered by two pairs of mirrors tilting in orthogonal directions on a MEMS device to address any qubit in a chain (steered beam is shown in green). The trap axial axis is rotated by 45\si{\degree} with respect to both tilting axes of the MEMS mirrors to utilize orthogonal tilting mirrors in order to maximize the addressable qubits. The projection and beam combining optics are represented by a black box. \textbf{(c)} Energy level schematic of a $^{171}$Yb$^+$ ion. The red and blue lines indicate the two photon Raman transition for qubit operations. }
\end{figure}

The qubits are laser cooled to near the ground state of the radial motional modes and initialized by optical pumping at the start of the experiment. This is followed by the qubit and motional manipulations, driven by stimulated Raman transitions using the beat-note between two orthogonal beams generated from a mode-locked \SI{355}{\nano m} picosecond-pulsed laser \cite{hayes2010entanglement}. The qubits are then measured by state-dependent fluorescence. Photons scattered by each of the qubits are collected by a high numerical aperture lens (NA $\approx 0.6$) and coupled into individual multi-mode fibers in a fiber array and sent to separate detectors for individual qubit state detection, as shown in Fig. \ref{MEMS setup}(c) \cite{crain2019high}. Our scheme features negligible detection crosstalk at a level of $10^{-4}$ in state detection.

The optical setup for implementing Raman quantum gates is schematically illustrated in Fig. \ref{MEMS setup}(a) and (b). One of the two orthogonal Raman beams is a global beam with an elliptical profile that illuminates all of the qubits simultaneously. The optical power and beam waist radius of the global beam are \SI{40}{\milli \watt} and \SI{8}{\micro m} $\times$ \SI{110}{\micro m}. The other is a pair of tightly focused individual addressing beams which can be independently steered across the qubit chain using a MEMS device. Single-mode photonic crystal fibers are used to deliver the individual addressing beams to the beam-steering system and the global beam to beam-shaping optics \cite{colombe2014single}. We use acousto-optic modulators (AOMs) to control the frequency, phase and amplitude of all three beams prior to the fibers. Steering of each individual beam is accomplished by a pair of MEMS mirrors each tilting in orthogonal directions. The details of the beam steering system is described in Ref.~\cite{crain2014individual} and supplementary. 
A dichroic mirror is used to reflect the Raman beams and transmit the qubit state-dependent fluorescence. The combination of single mode fiber and MEMS mirrors lead to clean Gaussian beams and low intensity crosstalk on the neighboring qubits at the level of $4 \times 10^{-6}$ to $4 \times 10^{-5}$ with a beam waist radius of \SI{\sim 2.2}{\micro m} and an ion spacing of \SI{\sim 5}{\micro m}. The intensity crosstalk leads to a gate crosstalk, which is defined as the ratio of Rabi frequency between the target qubit and the neighboring qubit, at the level of $0.2\%$ to $0.6\%$. The total number of addressable qubits is \SI{\sim 11}, limited by the maximum tilting angle of the MEMS mirrors. A maximum optical power of \SI{10}{\milli W}, a safety limit to avoid degradation from the UV laser, is applied onto the MEMS mirror, which leads to \SI{1.5}{\milli W} addressing beams going into the chamber and a maximum Rabi frequency for the motional sideband transition of \SI{7}{\kilo Hz}.

\begin{figure*}[!htb]
\center{\includegraphics[width=0.98 \textwidth]{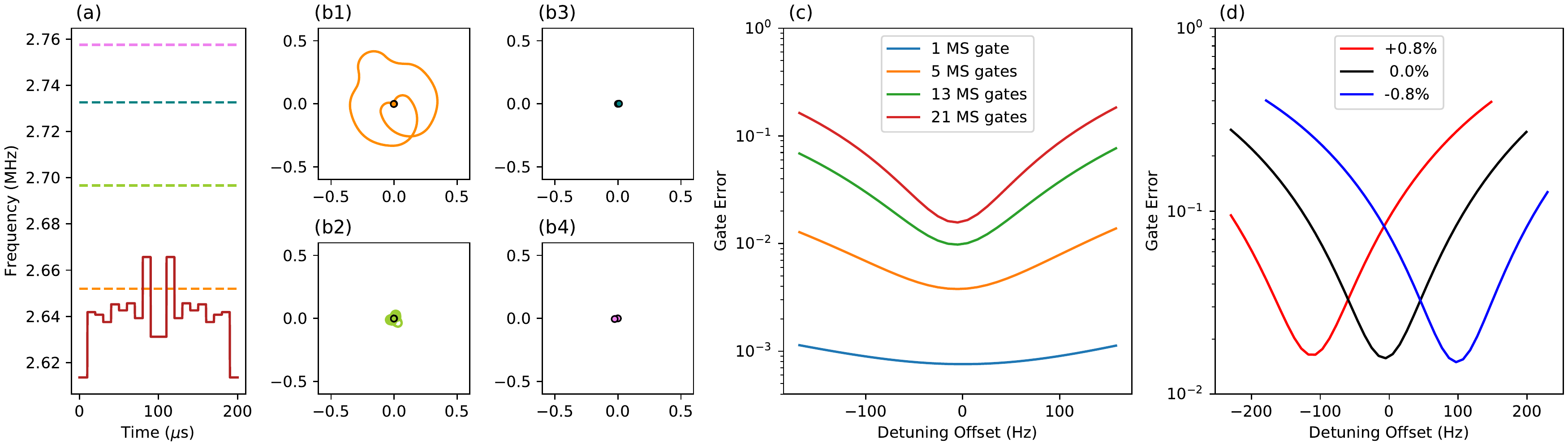}}\label{FM}
\caption{\label{FMgate}\textbf{(a)}Discrete frequency modulation pulse sequence in the present experiment. The solution consists of 20 symmetric segments. The total gate time is \SI{200}{\micro s}. The required sideband Rabi frequencies are \SI{5.55}{\kilo \hertz} and \SI{5.47}{\kilo \hertz} for FM gates in a 2-ion chain and a 4-ion chain. \textbf{(b)} Phase-space trajectory of four motional modes. \textbf{(c)} The estimated gross gate error of 1, 5, 13, 21 concatenated gates, given different detuning offsets. The estimation includes errors due to residual entanglement between spin and motion and the variation of the rotation angle. \textbf{(d)} The estimated of final-state error of 21 consecutive gates with $\pm0.8 \%$ deviation of Rabi frequency for the motional sideband transition. The amplitude error due to imperfect laser intensity can be compensated by intentional detuning offset.}
\end{figure*}

Robust FM MS gates, using a continuous waveform generated by arbitrary waveform generator (AWG), have been demonstrated in a $5$-qubit and $17$-qubit ion chains~\cite{leung2018robust,landsman2019two}. To be compatible with DDS, the scheme is reconstituted to its discrete analogue. The pulse is designed to be a sequence of equal-time segments, each of which has a constant frequency. The frequencies of the sequence are determined by a numerical optimizer, given the measured radial motional mode frequencies and a desired gate time as shown in Fig. \ref{FMgate}(a). The optimizer generates a pulse sequence which closes the phase-space trajectories of all radial motional modes and therefore disentangles the spins and the motion, as shown in Fig. \ref{FMgate}(b). It also constrains the Rabi frequency of the motional sideband transitions to be less than \SI{7}{\kilo Hz}.

The detuning error, arising from the drift of motional mode frequencies, leads to unwanted spin-motion entanglement and deviation of the geometric phase for the MS evolution. The error from residual entanglement can be made negligible over $\pm 1$ kHz detuning error ($< 2\times 10^{-5}$) in the robust FM gate~\cite{leung2018robust}. The accumulated phase deviation is represented by a deviation of the rotation angle of the gate, which can be considered as an amplitude error. In general amplitude errors are usually corrected for by tuning the laser intensity. However, if the detuning error changes on timescales faster than the time between calibration and the experimental circuit, then the intensity calibration is no longer accurate. Fig. \ref{FMgate}(c) shows the estimation of final-state fidelity after 1, 5, 13, and 21 consecutive MS gates are applied, as a function of detuning offset. The estimation considers both the residual spin-motion entanglement and the deviation of the rotation angle.
Taking advantage of the negligible residual spin-motion entanglement against detuning errors in robust FM gates, one can introduce intentional detuning offset to precisely compensate for the small amplitude error. As shown in Fig.\ref{FMgate}(d), a \SI{\pm 100}{Hz} detuning offset can compensate roughly $\mp 0.8 \%$ deviation of Rabi frequency for the motional sideband transition.

We designed an automatic calibration process for all critical parameters. The rough calibration is performed every 30 minutes and takes $\sim$10 minutes to complete. First, we calibrate the pointing accuracy of the two individual beams by tilting the MEMS mirror and observing the response of the target ion and neighboring ions to the beams. For each mirror, the tilt angle is calibrated by maximizing the population transfer of the target ion according to a single-qubit $\pi$ rotation pulse and minimizing the those of the neighboring ions after a single-qubit $10\pi$ rotation pulse. Next, we address a single ion in the chain and measure all of the mode frequencies by scanning the beat-note frequency and observing the motional sideband transition. The discrete FM solution is calculated based on the measured mode frequencies and the predetermined gate time of \SI{200}{\micro s}. After loading the resulting pulse sequence to the random access memory (RAM) of a field-programmable gate array (FPGA), the FPGA triggers the frequency updates for the DDS channels in real time during FM gates \cite{mount2016scalable}. The beam power is calibrated by ensuring the population of target ion in the $\ket{0}$ state to be 0.5 after an expected $3.5\pi$ single-qubit rotation. 

A final, fine calibration that takes tens of seconds is run just before the experiments to compensate for the small drift of the mode frequency and the laser intensity. As shown in Fig. \ref{FMgate}(d), the small drift of the laser intensity can be compensated by introducing a detuning offset, which is a more precise method than tuning the laser intensity. This calibration is done by scanning the detuning offset with 21 concatenated gates applied to $\ket{00}$. The detuning at which $\ket{00}$ and $\ket{11}$ have equal probability indicates the perfect rotation angle for the MS gate. On the average, this calibration 
improves our gate fidelity by about $0.5\%$ compare to just doing the rough calibration.

\begin{figure}[!htb]
\center{\includegraphics[width=0.48 \textwidth]{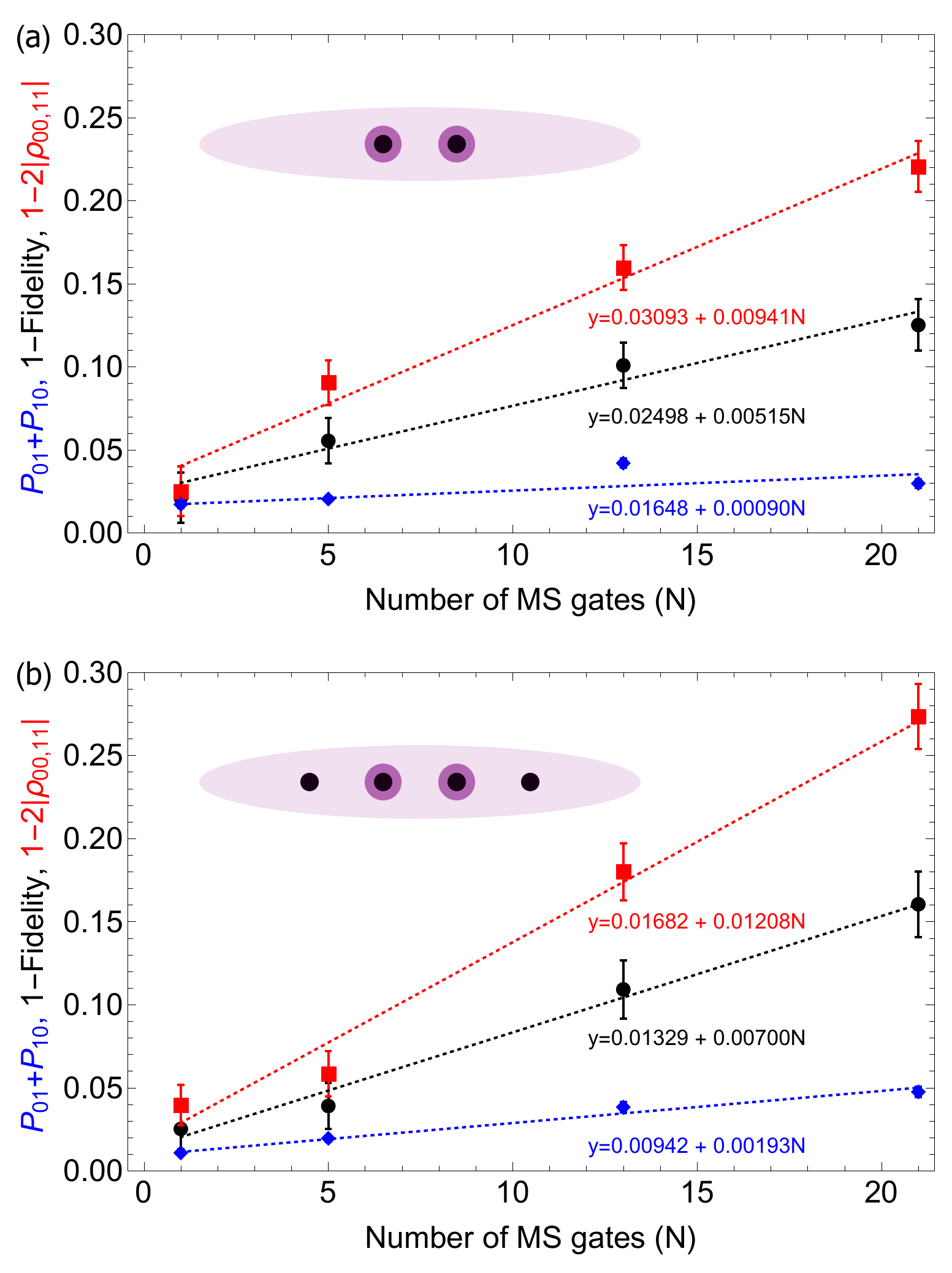}}
\caption{\label{fig:fidelity} Infidelity of the entangled state generated by repeated application of MS gates in a \textbf{(a)} two- and a \textbf{(b)} four-ion chain. The blue diamonds, red squares and black circles are the population leakage to $\ket{01}$ and $\ket{10}$ space, the loss of parity contrast, and the infidelity of the final state, respectively.
}
\end{figure}

We demonstrate the two-qubit MS gate in a two-ion chain and a four-ion chain. First, we initialize the target qubits to the $\ket{00}$ state and then apply a sequence of 1,5,13 and 21 MS gates to make the maximally entangled state $\ket{\psi_+} = (\ket{00} + i \ket{11})/\sqrt{2}$. We then extract the state fidelity by measuring the population and the parity contrast \cite{leibfried2003experimental}. The infidelity due to population leakage and the decrease of the parity contrast is plotted in the Fig. \ref{fig:fidelity}. The stochastic and the coherent error accumulate with concatenated gates in a linear and a quadratic way, respectively. However, the state preparation and measurement (SPAM) error remains constant. Using a linear fit for the data, we can extract the gate fidelity without the SPAM error. The two-qubit gate fidelity is $99.49(7)\%$ in a two-ion chain, and $99.30(6)\%$ in a four-ion chain. The data matches a linear fit, indicating that any coherent systematic error is negligible for the two-qubit gate in our system. The $\approx2\%$ SPAM error consists of $1.24\%$ preparation error and $0.98\%$ detection error according to the single-qubit gate gate-set-tomography (GST) analysis \cite{blume2017demonstration} performed on the present system. The high SPAM error is due to the limited control bandwidth on DDSs and can be suppressed to less than $0.1\%$ with an updated control system \cite{noek2013high}.

\begin{table*}[!htb]
\centering\begin{tabular}{c|cc}
    \hline
    Error source & \,\,\,\,\, Simulated error for 4-ion chain \,\,\,\,\, & \,\,\,\,\, Simulated error for 2-ion chain \\
    &($10^{-3}$)&($10^{-3}$)\\
    \hline
    Laser dephasing & $2.7\pm0.4$& $2.7\pm0.4$\\
    Motional dephasing & $1.2\pm0.1$ & $1.1\pm 0.1$\\
    Raman beam intensity fluctuation \,\,\,\, & 0.16 & 0.16\\
    Off-resonant coupling &$<0.3$ & $<0.3$\\
    Motional heating & 0.47 & 0.59\\
    Spontaneous emission & $<0.25$ & $<0.25$\\
    FM Solution imperfection & 0.76 & 0.04\\
    (due to laser power restriction) &&\\
    \hline
    Total & $5.84\pm0.5$ & $5.14\pm0.5$ \\
    \hline
  \end{tabular}
\caption{M\o lmer-S\o rensen gate Error budget. The errors are simulated with the full density matrix using the Master equation including various error sources. The laser and motional coherence times is measured to be $83.3\pm 11.5$ ms and $36.3 \pm 2.3$ ms, respectively. The beam intensity fluctuation is measured to be $0.8 \%$. The motional heating rate of the center-of-mass mode and the tilt mode in 2-ion chain is measured to be $\sim$200 phonons/s and $<10$ phonons/s, respectively. 
}
\end{table*}

To understand the residual error for the two-qubit gate, we study the impact of various error sources on an ideal two-qubit gate using numerical simulation  \cite{sorensen2000entanglement,ballance2016high,gaebler2016high}. The simulated error budget is shown in Table 1. Laser dephasing is the leading order effect. We use the optical phase-sensitive gate scheme \cite{InlekPRA2014} for our FM gates. This scheme 
leads to the sensitivities of the laser phase on our FM gates. We measure the laser coherence time with Ramsey interferometry using laser driven phase-sensitive single qubit operations. The Rabi frequency of this phase-sensitive operation is affected by motional states of radial modes, as describe by the Debye-Waller effect \cite{wineland1998experimental,sorensen2000entanglement}. During the waiting time of Ramsey interferometry, the motional state is heated from near ground state to a thermal state due to anomalous heating~\cite{DeslauriersPRL2006,HitePRL2012}. The heating rate of the center-of-mass mode and the tilt mode in the 2-ion chain is measured to be $\sim$200 phonons/s and $<10$ phonons/s, respectively. Therefore, the Ramsey contrast should be amended with the Debye-Waller effect. The corrected Ramsey contrast lead to laser coherence time of $83.3\pm 11.5$ ms. On the other hand, the qubit coherence time is close to 1 second if we use microwave or laser driven phase-insensitive single-qubit operations for Ramsey interferometry. This significant reduction from 1 second to $83.3\pm 11.5$ ms is caused by optical-phase fluctuations of two Raman beams at the qubit location arising from the variations of the optical path length.

Motional dephasing is the next significant source, and has many potential mechanisms \cite{wineland1998experimental}. In our system, it is mainly due to the amplitude fluctuation of the RF source used to generate the trapping potential. We apply a Ramsey interferometry to the motional sideband transition to measure the motional coherence time. To avoid the Debye-Waller effect, the measurement is done on the zig-zag mode of a 7-ion chain, which features negligible anomalous heating. The motional coherence time is measured to be $36.3 \pm 2.3$ ms. 

The intensity fluctuation of the tightly focused addressing Raman beams of $<0.8\%$ is deduced by observing the decay of Rabi flopping for a phase-insensitive single qubit gate, driven by a co-propagating pair of Raman beams. The intensity fluctuation of the global beam should be at the same level. The upper bound of the off-resonant coupling to the carrier transition is estimated using the equation in Ref. \cite{sorensen2000entanglement}.

The dominant error sources in our scheme are entirely technical in nature. The fluctuation of the Raman beam path and intensity can be suppressed by better optomechanical design. The noise from the RF source can be suppressed by active feedback on the RF amplitude and better mechanical stability of the helical resonator.We observe UV-induced damage on the MEMS mirrors when the Raman beam power is increased to above \SI{10}{\milli W}, so we increase the optical power of the global beam to maintain gate speeds under such power constraint on the individual beams. A faster gate will significantly suppress the error from laser and motional dephasing. 
The error due to spontaneous emission can be suppressed to be less than $1.7 \times 10^{-4}$ by balancing global and addressing beam intensities.
With the achievable laser coherence time ($\sim$1 s), motional coherence time ($\sim 0.5$ s) \cite{ballance2016high}, and negligible spontaneous emission rate 
\cite{ozeri2007errors}, a two-qubit gate with fidelities well over $99.9\%$ is possible in a long ion chain.

\acknowledgements
This work was primarily supported by the Office of the Director of National Intelligence - Intelligence Advanced Research Projects Activity through ARO contract W911NF-16- 1-0082 (experimental implementation), by DOE BES award de-sc0019449 (error analysis), by National Science Foundation Expeditions in Computing award 1730104 (FM gate design), by National Science Foundation STAQ project Phy-1818914 (simulation studies), and by DOE ASCR award de-sc0019294 (measurement protocol).

\nocite{supplemental1}
\nocite{hayes2012remote}
\nocite{lindblad1976generators}
\nocite{gardiner2004quantum}

\bibliographystyle{apsrev}
\bibliography{References}

\begin{thebibliography}{47}
\expandafter\ifx\csname natexlab\endcsname\relax\def\natexlab#1{#1}\fi
\expandafter\ifx\csname bibnamefont\endcsname\relax
  \def\bibnamefont#1{#1}\fi
\expandafter\ifx\csname bibfnamefont\endcsname\relax
  \def\bibfnamefont#1{#1}\fi
\expandafter\ifx\csname citenamefont\endcsname\relax
  \def\citenamefont#1{#1}\fi
\expandafter\ifx\csname url\endcsname\relax
  \def\url#1{\texttt{#1}}\fi
\expandafter\ifx\csname urlprefix\endcsname\relax\def\urlprefix{URL }\fi
\providecommand{\bibinfo}[2]{#2}
\providecommand{\eprint}[2][]{\url{#2}}

\bibitem[{\citenamefont{Wang et~al.}(2017)\citenamefont{Wang, Um, Zhang, An,
  Lyu, Zhang, Duan, Yum, and Kim}}]{wang2017single}
\bibinfo{author}{\bibfnamefont{Y.}~\bibnamefont{Wang}},
  \bibinfo{author}{\bibfnamefont{M.}~\bibnamefont{Um}},
  \bibinfo{author}{\bibfnamefont{J.}~\bibnamefont{Zhang}},
  \bibinfo{author}{\bibfnamefont{S.}~\bibnamefont{An}},
  \bibinfo{author}{\bibfnamefont{M.}~\bibnamefont{Lyu}},
  \bibinfo{author}{\bibfnamefont{J.-N.} \bibnamefont{Zhang}},
  \bibinfo{author}{\bibfnamefont{L.-M.} \bibnamefont{Duan}},
  \bibinfo{author}{\bibfnamefont{D.}~\bibnamefont{Yum}}, \bibnamefont{and}
  \bibinfo{author}{\bibfnamefont{K.}~\bibnamefont{Kim}},
  \bibinfo{journal}{Nature Photonics} \textbf{\bibinfo{volume}{11}},
  \bibinfo{pages}{646} (\bibinfo{year}{2017}).

\bibitem[{\citenamefont{Crain et~al.}(2019)\citenamefont{Crain, Cahall,
  Vrijsen, Wollman, Shaw, Verma, Nam, and Kim}}]{crain2019high}
\bibinfo{author}{\bibfnamefont{S.}~\bibnamefont{Crain}},
  \bibinfo{author}{\bibfnamefont{C.}~\bibnamefont{Cahall}},
  \bibinfo{author}{\bibfnamefont{G.}~\bibnamefont{Vrijsen}},
  \bibinfo{author}{\bibfnamefont{E.~E.} \bibnamefont{Wollman}},
  \bibinfo{author}{\bibfnamefont{M.~D.} \bibnamefont{Shaw}},
  \bibinfo{author}{\bibfnamefont{V.~B.} \bibnamefont{Verma}},
  \bibinfo{author}{\bibfnamefont{S.~W.} \bibnamefont{Nam}}, \bibnamefont{and}
  \bibinfo{author}{\bibfnamefont{J.}~\bibnamefont{Kim}},
  \bibinfo{journal}{Communications Physics} \textbf{\bibinfo{volume}{2}},
  \bibinfo{pages}{1} (\bibinfo{year}{2019}).

\bibitem[{\citenamefont{Ballance et~al.}(2016)\citenamefont{Ballance, Harty,
  Linke, Sepiol, and Lucas}}]{ballance2016high}
\bibinfo{author}{\bibfnamefont{C.}~\bibnamefont{Ballance}},
  \bibinfo{author}{\bibfnamefont{T.}~\bibnamefont{Harty}},
  \bibinfo{author}{\bibfnamefont{N.}~\bibnamefont{Linke}},
  \bibinfo{author}{\bibfnamefont{M.}~\bibnamefont{Sepiol}}, \bibnamefont{and}
  \bibinfo{author}{\bibfnamefont{D.}~\bibnamefont{Lucas}},
  \bibinfo{journal}{Physical Review Letters} \textbf{\bibinfo{volume}{117}},
  \bibinfo{pages}{060504} (\bibinfo{year}{2016}).

\bibitem[{\citenamefont{Gaebler et~al.}(2016)\citenamefont{Gaebler, Tan, Lin,
  Wan, Bowler, Keith, Glancy, Coakley, Knill, Leibfried
  et~al.}}]{gaebler2016high}
\bibinfo{author}{\bibfnamefont{J.~P.} \bibnamefont{Gaebler}},
  \bibinfo{author}{\bibfnamefont{T.~R.} \bibnamefont{Tan}},
  \bibinfo{author}{\bibfnamefont{Y.}~\bibnamefont{Lin}},
  \bibinfo{author}{\bibfnamefont{Y.}~\bibnamefont{Wan}},
  \bibinfo{author}{\bibfnamefont{R.}~\bibnamefont{Bowler}},
  \bibinfo{author}{\bibfnamefont{A.~C.} \bibnamefont{Keith}},
  \bibinfo{author}{\bibfnamefont{S.}~\bibnamefont{Glancy}},
  \bibinfo{author}{\bibfnamefont{K.}~\bibnamefont{Coakley}},
  \bibinfo{author}{\bibfnamefont{E.}~\bibnamefont{Knill}},
  \bibinfo{author}{\bibfnamefont{D.}~\bibnamefont{Leibfried}},
  \bibnamefont{et~al.}, \bibinfo{journal}{Physical Review Letters}
  \textbf{\bibinfo{volume}{117}}, \bibinfo{pages}{060505}
  (\bibinfo{year}{2016}).

\bibitem[{\citenamefont{Harty et~al.}(2014)\citenamefont{Harty, Allcock,
  Ballance, Guidoni, Janacek, Linke, Stacey, and Lucas}}]{harty2014high}
\bibinfo{author}{\bibfnamefont{T.}~\bibnamefont{Harty}},
  \bibinfo{author}{\bibfnamefont{D.}~\bibnamefont{Allcock}},
  \bibinfo{author}{\bibfnamefont{C.~J.} \bibnamefont{Ballance}},
  \bibinfo{author}{\bibfnamefont{L.}~\bibnamefont{Guidoni}},
  \bibinfo{author}{\bibfnamefont{H.}~\bibnamefont{Janacek}},
  \bibinfo{author}{\bibfnamefont{N.}~\bibnamefont{Linke}},
  \bibinfo{author}{\bibfnamefont{D.}~\bibnamefont{Stacey}}, \bibnamefont{and}
  \bibinfo{author}{\bibfnamefont{D.}~\bibnamefont{Lucas}},
  \bibinfo{journal}{Physical Review Letters} \textbf{\bibinfo{volume}{113}},
  \bibinfo{pages}{220501} (\bibinfo{year}{2014}).

\bibitem[{\citenamefont{Pino et~al.}(2020)\citenamefont{Pino, Dreiling,
  Figgatt, Gaebler, Moses, Baldwin, Foss-Feig, Hayes, Mayer, Ryan-Anderson
  et~al.}}]{pino2020demonstration}
\bibinfo{author}{\bibfnamefont{J.}~\bibnamefont{Pino}},
  \bibinfo{author}{\bibfnamefont{J.}~\bibnamefont{Dreiling}},
  \bibinfo{author}{\bibfnamefont{C.}~\bibnamefont{Figgatt}},
  \bibinfo{author}{\bibfnamefont{J.}~\bibnamefont{Gaebler}},
  \bibinfo{author}{\bibfnamefont{S.}~\bibnamefont{Moses}},
  \bibinfo{author}{\bibfnamefont{C.}~\bibnamefont{Baldwin}},
  \bibinfo{author}{\bibfnamefont{M.}~\bibnamefont{Foss-Feig}},
  \bibinfo{author}{\bibfnamefont{D.}~\bibnamefont{Hayes}},
  \bibinfo{author}{\bibfnamefont{K.}~\bibnamefont{Mayer}},
  \bibinfo{author}{\bibfnamefont{C.}~\bibnamefont{Ryan-Anderson}},
  \bibnamefont{et~al.}, \bibinfo{journal}{arXiv preprint arXiv:2003.01293}
  (\bibinfo{year}{2020}).

\bibitem[{\citenamefont{S{\o}rensen and M{\o}lmer}(1999)}]{sorensen1999quantum}
\bibinfo{author}{\bibfnamefont{A.}~\bibnamefont{S{\o}rensen}} \bibnamefont{and}
  \bibinfo{author}{\bibfnamefont{K.}~\bibnamefont{M{\o}lmer}},
  \bibinfo{journal}{Physical Review Letters} \textbf{\bibinfo{volume}{82}},
  \bibinfo{pages}{1971} (\bibinfo{year}{1999}).

\bibitem[{\citenamefont{S{\o}rensen and
  M{\o}lmer}(2000)}]{sorensen2000entanglement}
\bibinfo{author}{\bibfnamefont{A.}~\bibnamefont{S{\o}rensen}} \bibnamefont{and}
  \bibinfo{author}{\bibfnamefont{K.}~\bibnamefont{M{\o}lmer}},
  \bibinfo{journal}{Physical Review A} \textbf{\bibinfo{volume}{62}},
  \bibinfo{pages}{022311} (\bibinfo{year}{2000}).

\bibitem[{\citenamefont{Lanyon et~al.}(2011)\citenamefont{Lanyon, Hempel, Nigg,
  M{\"u}ller, Gerritsma, Z{\"a}hringer, Schindler, Barreiro, Rambach, Kirchmair
  et~al.}}]{lanyon2011universal}
\bibinfo{author}{\bibfnamefont{B.~P.} \bibnamefont{Lanyon}},
  \bibinfo{author}{\bibfnamefont{C.}~\bibnamefont{Hempel}},
  \bibinfo{author}{\bibfnamefont{D.}~\bibnamefont{Nigg}},
  \bibinfo{author}{\bibfnamefont{M.}~\bibnamefont{M{\"u}ller}},
  \bibinfo{author}{\bibfnamefont{R.}~\bibnamefont{Gerritsma}},
  \bibinfo{author}{\bibfnamefont{F.}~\bibnamefont{Z{\"a}hringer}},
  \bibinfo{author}{\bibfnamefont{P.}~\bibnamefont{Schindler}},
  \bibinfo{author}{\bibfnamefont{J.~T.} \bibnamefont{Barreiro}},
  \bibinfo{author}{\bibfnamefont{M.}~\bibnamefont{Rambach}},
  \bibinfo{author}{\bibfnamefont{G.}~\bibnamefont{Kirchmair}},
  \bibnamefont{et~al.}, \bibinfo{journal}{Science}
  \textbf{\bibinfo{volume}{334}}, \bibinfo{pages}{57} (\bibinfo{year}{2011}).

\bibitem[{\citenamefont{Shor}(1995)}]{shor1995scheme}
\bibinfo{author}{\bibfnamefont{P.~W.} \bibnamefont{Shor}},
  \bibinfo{journal}{Physical Review A} \textbf{\bibinfo{volume}{52}},
  \bibinfo{pages}{R2493} (\bibinfo{year}{1995}).

\bibitem[{\citenamefont{Kitaev}(2003)}]{kitaev2003fault}
\bibinfo{author}{\bibfnamefont{A.~Y.} \bibnamefont{Kitaev}},
  \bibinfo{journal}{Annals of Physics} \textbf{\bibinfo{volume}{303}},
  \bibinfo{pages}{2} (\bibinfo{year}{2003}).

\bibitem[{\citenamefont{Aharonov and Ben-Or}(2008)}]{aharonov1999fault}
\bibinfo{author}{\bibfnamefont{D.}~\bibnamefont{Aharonov}} \bibnamefont{and}
  \bibinfo{author}{\bibfnamefont{M.}~\bibnamefont{Ben-Or}},
  \bibinfo{journal}{SIAM Journal on Computing}  (\bibinfo{year}{2008}).

\bibitem[{\citenamefont{Debnath et~al.}(2016)\citenamefont{Debnath, Linke,
  Figgatt, Landsman, Wright, and Monroe}}]{debnath2016demonstration}
\bibinfo{author}{\bibfnamefont{S.}~\bibnamefont{Debnath}},
  \bibinfo{author}{\bibfnamefont{N.~M.} \bibnamefont{Linke}},
  \bibinfo{author}{\bibfnamefont{C.}~\bibnamefont{Figgatt}},
  \bibinfo{author}{\bibfnamefont{K.~A.} \bibnamefont{Landsman}},
  \bibinfo{author}{\bibfnamefont{K.}~\bibnamefont{Wright}}, \bibnamefont{and}
  \bibinfo{author}{\bibfnamefont{C.}~\bibnamefont{Monroe}},
  \bibinfo{journal}{Nature} \textbf{\bibinfo{volume}{536}}, \bibinfo{pages}{63}
  (\bibinfo{year}{2016}).

\bibitem[{\citenamefont{Wright et~al.}(2019)\citenamefont{Wright, Beck,
  Debnath, Amini, Nam, Grzesiak, Chen, Pisenti, Chmielewski, Collins
  et~al.}}]{wright2019benchmarking}
\bibinfo{author}{\bibfnamefont{K.}~\bibnamefont{Wright}},
  \bibinfo{author}{\bibfnamefont{K.}~\bibnamefont{Beck}},
  \bibinfo{author}{\bibfnamefont{S.}~\bibnamefont{Debnath}},
  \bibinfo{author}{\bibfnamefont{J.}~\bibnamefont{Amini}},
  \bibinfo{author}{\bibfnamefont{Y.}~\bibnamefont{Nam}},
  \bibinfo{author}{\bibfnamefont{N.}~\bibnamefont{Grzesiak}},
  \bibinfo{author}{\bibfnamefont{J.-S.} \bibnamefont{Chen}},
  \bibinfo{author}{\bibfnamefont{N.}~\bibnamefont{Pisenti}},
  \bibinfo{author}{\bibfnamefont{M.}~\bibnamefont{Chmielewski}},
  \bibinfo{author}{\bibfnamefont{C.}~\bibnamefont{Collins}},
  \bibnamefont{et~al.}, \bibinfo{journal}{Nature Communications}
  \textbf{\bibinfo{volume}{10}}, \bibinfo{pages}{1} (\bibinfo{year}{2019}).

\bibitem[{\citenamefont{Schmidt-Kaler et~al.}(2003)\citenamefont{Schmidt-Kaler,
  H{\"a}ffner, Riebe, Gulde, Lancaster, Deuschle, Becher, Roos, Eschner, and
  Blatt}}]{schmidt2003nature}
\bibinfo{author}{\bibfnamefont{F.}~\bibnamefont{Schmidt-Kaler}},
  \bibinfo{author}{\bibfnamefont{H.}~\bibnamefont{H{\"a}ffner}},
  \bibinfo{author}{\bibfnamefont{M.}~\bibnamefont{Riebe}},
  \bibinfo{author}{\bibfnamefont{S.}~\bibnamefont{Gulde}},
  \bibinfo{author}{\bibfnamefont{G.}~\bibnamefont{Lancaster}},
  \bibinfo{author}{\bibfnamefont{T.}~\bibnamefont{Deuschle}},
  \bibinfo{author}{\bibfnamefont{C.}~\bibnamefont{Becher}},
  \bibinfo{author}{\bibfnamefont{C.}~\bibnamefont{Roos}},
  \bibinfo{author}{\bibfnamefont{J.}~\bibnamefont{Eschner}}, \bibnamefont{and}
  \bibinfo{author}{\bibfnamefont{R.}~\bibnamefont{Blatt}},
  \bibinfo{journal}{Nature} \textbf{\bibinfo{volume}{422}},
  \bibinfo{pages}{412} (\bibinfo{year}{2003}).

\bibitem[{\citenamefont{Yavuz et~al.}(2006)\citenamefont{Yavuz, Kulatunga,
  Urban, Johnson, Proite, Henage, Walker, and Saffman}}]{yavuz2006fast}
\bibinfo{author}{\bibfnamefont{D.}~\bibnamefont{Yavuz}},
  \bibinfo{author}{\bibfnamefont{P.}~\bibnamefont{Kulatunga}},
  \bibinfo{author}{\bibfnamefont{E.}~\bibnamefont{Urban}},
  \bibinfo{author}{\bibfnamefont{T.~A.} \bibnamefont{Johnson}},
  \bibinfo{author}{\bibfnamefont{N.}~\bibnamefont{Proite}},
  \bibinfo{author}{\bibfnamefont{T.}~\bibnamefont{Henage}},
  \bibinfo{author}{\bibfnamefont{T.}~\bibnamefont{Walker}}, \bibnamefont{and}
  \bibinfo{author}{\bibfnamefont{M.}~\bibnamefont{Saffman}},
  \bibinfo{journal}{Physical Review Letters} \textbf{\bibinfo{volume}{96}},
  \bibinfo{pages}{063001} (\bibinfo{year}{2006}).

\bibitem[{\citenamefont{Knoernschild et~al.}(2010)\citenamefont{Knoernschild,
  Zhang, Isenhower, Gill, Lu, Saffman, and Kim}}]{KnoernschildAPL2010}
\bibinfo{author}{\bibfnamefont{C.}~\bibnamefont{Knoernschild}},
  \bibinfo{author}{\bibfnamefont{X.~L.} \bibnamefont{Zhang}},
  \bibinfo{author}{\bibfnamefont{L.}~\bibnamefont{Isenhower}},
  \bibinfo{author}{\bibfnamefont{A.~T.} \bibnamefont{Gill}},
  \bibinfo{author}{\bibfnamefont{F.~P.} \bibnamefont{Lu}},
  \bibinfo{author}{\bibfnamefont{M.}~\bibnamefont{Saffman}}, \bibnamefont{and}
  \bibinfo{author}{\bibfnamefont{J.}~\bibnamefont{Kim}},
  \bibinfo{journal}{Appl. Phys. Lett.} \textbf{\bibinfo{volume}{97}},
  \bibinfo{pages}{134101} (\bibinfo{year}{2010}).

\bibitem[{\citenamefont{Crain et~al.}(2014)\citenamefont{Crain, Mount, Baek,
  and Kim}}]{crain2014individual}
\bibinfo{author}{\bibfnamefont{S.}~\bibnamefont{Crain}},
  \bibinfo{author}{\bibfnamefont{E.}~\bibnamefont{Mount}},
  \bibinfo{author}{\bibfnamefont{S.}~\bibnamefont{Baek}}, \bibnamefont{and}
  \bibinfo{author}{\bibfnamefont{J.}~\bibnamefont{Kim}},
  \bibinfo{journal}{Applied Physics Letters} \textbf{\bibinfo{volume}{105}},
  \bibinfo{pages}{181115} (\bibinfo{year}{2014}).

\bibitem[{\citenamefont{Lee et~al.}(2005)\citenamefont{Lee, Brickman,
  Deslauriers, Haljan, Duan, and Monroe}}]{LeeJOB2005}
\bibinfo{author}{\bibfnamefont{P.~J.} \bibnamefont{Lee}},
  \bibinfo{author}{\bibfnamefont{K.-A.} \bibnamefont{Brickman}},
  \bibinfo{author}{\bibfnamefont{L.}~\bibnamefont{Deslauriers}},
  \bibinfo{author}{\bibfnamefont{P.~C.} \bibnamefont{Haljan}},
  \bibinfo{author}{\bibfnamefont{L.-M.} \bibnamefont{Duan}}, \bibnamefont{and}
  \bibinfo{author}{\bibfnamefont{C.}~\bibnamefont{Monroe}},
  \bibinfo{journal}{Journal of Optics B: Quantum and Semiclassical Optics}
  \textbf{\bibinfo{volume}{7}}, \bibinfo{pages}{S371} (\bibinfo{year}{2005}).

\bibitem[{\citenamefont{Inlek et~al.}(2014)\citenamefont{Inlek, Vittorini,
  Hucul, Crocker, and Monroe}}]{InlekPRA2014}
\bibinfo{author}{\bibfnamefont{I.~V.} \bibnamefont{Inlek}},
  \bibinfo{author}{\bibfnamefont{G.}~\bibnamefont{Vittorini}},
  \bibinfo{author}{\bibfnamefont{D.}~\bibnamefont{Hucul}},
  \bibinfo{author}{\bibfnamefont{C.}~\bibnamefont{Crocker}}, \bibnamefont{and}
  \bibinfo{author}{\bibfnamefont{C.}~\bibnamefont{Monroe}},
  \bibinfo{journal}{Phys. Rev. A} \textbf{\bibinfo{volume}{90}},
  \bibinfo{pages}{042316} (\bibinfo{year}{2014}).

\bibitem[{\citenamefont{Tan et~al.}(2015)\citenamefont{Tan, Gaebler, Lin, Wan,
  Bowler, Leibfried, and Wineland}}]{TanNature2015}
\bibinfo{author}{\bibfnamefont{T.~R.} \bibnamefont{Tan}},
  \bibinfo{author}{\bibfnamefont{J.~P.} \bibnamefont{Gaebler}},
  \bibinfo{author}{\bibfnamefont{Y.}~\bibnamefont{Lin}},
  \bibinfo{author}{\bibfnamefont{Y.}~\bibnamefont{Wan}},
  \bibinfo{author}{\bibfnamefont{R.}~\bibnamefont{Bowler}},
  \bibinfo{author}{\bibfnamefont{D.}~\bibnamefont{Leibfried}},
  \bibnamefont{and} \bibinfo{author}{\bibfnamefont{D.~J.}
  \bibnamefont{Wineland}}, \bibinfo{journal}{Nature}
  \textbf{\bibinfo{volume}{528}}, \bibinfo{pages}{380} (\bibinfo{year}{2015}).

\bibitem[{\citenamefont{Zhu et~al.}(2006)\citenamefont{Zhu, Monroe, and
  Duan}}]{zhu2006arbitrary}
\bibinfo{author}{\bibfnamefont{S.-L.} \bibnamefont{Zhu}},
  \bibinfo{author}{\bibfnamefont{C.}~\bibnamefont{Monroe}}, \bibnamefont{and}
  \bibinfo{author}{\bibfnamefont{L.-M.} \bibnamefont{Duan}},
  \bibinfo{journal}{EPL (Europhysics Letters)} \textbf{\bibinfo{volume}{73}},
  \bibinfo{pages}{485} (\bibinfo{year}{2006}).

\bibitem[{\citenamefont{Roos}(2008)}]{roos2008ion}
\bibinfo{author}{\bibfnamefont{C.~F.} \bibnamefont{Roos}},
  \bibinfo{journal}{New Journal of Physics} \textbf{\bibinfo{volume}{10}},
  \bibinfo{pages}{013002} (\bibinfo{year}{2008}).

\bibitem[{\citenamefont{Green and Biercuk}(2015)}]{green2015phase}
\bibinfo{author}{\bibfnamefont{T.~J.} \bibnamefont{Green}} \bibnamefont{and}
  \bibinfo{author}{\bibfnamefont{M.~J.} \bibnamefont{Biercuk}},
  \bibinfo{journal}{Physical Review Letters} \textbf{\bibinfo{volume}{114}},
  \bibinfo{pages}{120502} (\bibinfo{year}{2015}).

\bibitem[{\citenamefont{Milne et~al.}(2020)\citenamefont{Milne, Edmunds,
  Hempel, Roy, Mavadia, and Biercuk}}]{milne2020phase}
\bibinfo{author}{\bibfnamefont{A.~R.} \bibnamefont{Milne}},
  \bibinfo{author}{\bibfnamefont{C.~L.} \bibnamefont{Edmunds}},
  \bibinfo{author}{\bibfnamefont{C.}~\bibnamefont{Hempel}},
  \bibinfo{author}{\bibfnamefont{F.}~\bibnamefont{Roy}},
  \bibinfo{author}{\bibfnamefont{S.}~\bibnamefont{Mavadia}}, \bibnamefont{and}
  \bibinfo{author}{\bibfnamefont{M.~J.} \bibnamefont{Biercuk}},
  \bibinfo{journal}{Physical Review Applied} \textbf{\bibinfo{volume}{13}},
  \bibinfo{pages}{024022} (\bibinfo{year}{2020}).

\bibitem[{\citenamefont{Lu et~al.}(2019)\citenamefont{Lu, Zhang, Zhang, Chen,
  Shen, Zhang, Zhang, and Kim}}]{lu2019global}
\bibinfo{author}{\bibfnamefont{Y.}~\bibnamefont{Lu}},
  \bibinfo{author}{\bibfnamefont{S.}~\bibnamefont{Zhang}},
  \bibinfo{author}{\bibfnamefont{K.}~\bibnamefont{Zhang}},
  \bibinfo{author}{\bibfnamefont{W.}~\bibnamefont{Chen}},
  \bibinfo{author}{\bibfnamefont{Y.}~\bibnamefont{Shen}},
  \bibinfo{author}{\bibfnamefont{J.}~\bibnamefont{Zhang}},
  \bibinfo{author}{\bibfnamefont{J.-N.} \bibnamefont{Zhang}}, \bibnamefont{and}
  \bibinfo{author}{\bibfnamefont{K.}~\bibnamefont{Kim}},
  \bibinfo{journal}{Nature} \textbf{\bibinfo{volume}{572}},
  \bibinfo{pages}{363} (\bibinfo{year}{2019}).

\bibitem[{\citenamefont{Shapira et~al.}(2018)\citenamefont{Shapira, Shaniv,
  Manovitz, Akerman, and Ozeri}}]{shapira2018robust}
\bibinfo{author}{\bibfnamefont{Y.}~\bibnamefont{Shapira}},
  \bibinfo{author}{\bibfnamefont{R.}~\bibnamefont{Shaniv}},
  \bibinfo{author}{\bibfnamefont{T.}~\bibnamefont{Manovitz}},
  \bibinfo{author}{\bibfnamefont{N.}~\bibnamefont{Akerman}}, \bibnamefont{and}
  \bibinfo{author}{\bibfnamefont{R.}~\bibnamefont{Ozeri}},
  \bibinfo{journal}{Physical Review Letters} \textbf{\bibinfo{volume}{121}},
  \bibinfo{pages}{180502} (\bibinfo{year}{2018}).

\bibitem[{\citenamefont{Blumel et~al.}(2019)\citenamefont{Blumel, Grzesiak, and
  Nam}}]{blumel2019power}
\bibinfo{author}{\bibfnamefont{R.}~\bibnamefont{Blumel}},
  \bibinfo{author}{\bibfnamefont{N.}~\bibnamefont{Grzesiak}}, \bibnamefont{and}
  \bibinfo{author}{\bibfnamefont{Y.}~\bibnamefont{Nam}},
  \bibinfo{journal}{arXiv preprint arXiv:1905.09292}  (\bibinfo{year}{2019}).

\bibitem[{\citenamefont{Leung et~al.}(2018)\citenamefont{Leung, Landsman,
  Figgatt, Linke, Monroe, and Brown}}]{leung2018robust}
\bibinfo{author}{\bibfnamefont{P.~H.} \bibnamefont{Leung}},
  \bibinfo{author}{\bibfnamefont{K.~A.} \bibnamefont{Landsman}},
  \bibinfo{author}{\bibfnamefont{C.}~\bibnamefont{Figgatt}},
  \bibinfo{author}{\bibfnamefont{N.~M.} \bibnamefont{Linke}},
  \bibinfo{author}{\bibfnamefont{C.}~\bibnamefont{Monroe}}, \bibnamefont{and}
  \bibinfo{author}{\bibfnamefont{K.~R.} \bibnamefont{Brown}},
  \bibinfo{journal}{Physical Review Letters} \textbf{\bibinfo{volume}{120}},
  \bibinfo{pages}{020501} (\bibinfo{year}{2018}).

\bibitem[{\citenamefont{Leung and Brown}(2018)}]{leung2018arbitrary}
\bibinfo{author}{\bibfnamefont{P.~H.} \bibnamefont{Leung}} \bibnamefont{and}
  \bibinfo{author}{\bibfnamefont{K.~R.} \bibnamefont{Brown}},
  \bibinfo{journal}{Physical Review A} \textbf{\bibinfo{volume}{98}},
  \bibinfo{pages}{032318} (\bibinfo{year}{2018}).

\bibitem[{\citenamefont{Landsman et~al.}(2019)\citenamefont{Landsman, Wu,
  Leung, Zhu, Linke, Brown, Duan, and Monroe}}]{landsman2019two}
\bibinfo{author}{\bibfnamefont{K.~A.} \bibnamefont{Landsman}},
  \bibinfo{author}{\bibfnamefont{Y.}~\bibnamefont{Wu}},
  \bibinfo{author}{\bibfnamefont{P.~H.} \bibnamefont{Leung}},
  \bibinfo{author}{\bibfnamefont{D.}~\bibnamefont{Zhu}},
  \bibinfo{author}{\bibfnamefont{N.~M.} \bibnamefont{Linke}},
  \bibinfo{author}{\bibfnamefont{K.~R.} \bibnamefont{Brown}},
  \bibinfo{author}{\bibfnamefont{L.}~\bibnamefont{Duan}}, \bibnamefont{and}
  \bibinfo{author}{\bibfnamefont{C.}~\bibnamefont{Monroe}},
  \bibinfo{journal}{Physical Review A} \textbf{\bibinfo{volume}{100}},
  \bibinfo{pages}{022332} (\bibinfo{year}{2019}).

\bibitem[{\citenamefont{Olmschenk et~al.}(2007)\citenamefont{Olmschenk, Younge,
  Moehring, Matsukevich, Maunz, and Monroe}}]{olmschenk2007manipulation}
\bibinfo{author}{\bibfnamefont{S.}~\bibnamefont{Olmschenk}},
  \bibinfo{author}{\bibfnamefont{K.~C.} \bibnamefont{Younge}},
  \bibinfo{author}{\bibfnamefont{D.~L.} \bibnamefont{Moehring}},
  \bibinfo{author}{\bibfnamefont{D.~N.} \bibnamefont{Matsukevich}},
  \bibinfo{author}{\bibfnamefont{P.}~\bibnamefont{Maunz}}, \bibnamefont{and}
  \bibinfo{author}{\bibfnamefont{C.}~\bibnamefont{Monroe}},
  \bibinfo{journal}{Physical Review A} \textbf{\bibinfo{volume}{76}},
  \bibinfo{pages}{052314} (\bibinfo{year}{2007}).

\bibitem[{\citenamefont{Maunz}(2016)}]{maunz2016high}
\bibinfo{author}{\bibfnamefont{P.~L.~W.} \bibnamefont{Maunz}},
  \bibinfo{type}{Tech. Rep.}, \bibinfo{institution}{Sandia National
  Lab.(SNL-NM), Albuquerque, NM (United States)} (\bibinfo{year}{2016}).

\bibitem[{\citenamefont{Hayes et~al.}(2010)\citenamefont{Hayes, Matsukevich,
  Maunz, Hucul, Quraishi, Olmschenk, Campbell, Mizrahi, Senko, and
  Monroe}}]{hayes2010entanglement}
\bibinfo{author}{\bibfnamefont{D.}~\bibnamefont{Hayes}},
  \bibinfo{author}{\bibfnamefont{D.~N.} \bibnamefont{Matsukevich}},
  \bibinfo{author}{\bibfnamefont{P.}~\bibnamefont{Maunz}},
  \bibinfo{author}{\bibfnamefont{D.}~\bibnamefont{Hucul}},
  \bibinfo{author}{\bibfnamefont{Q.}~\bibnamefont{Quraishi}},
  \bibinfo{author}{\bibfnamefont{S.}~\bibnamefont{Olmschenk}},
  \bibinfo{author}{\bibfnamefont{W.}~\bibnamefont{Campbell}},
  \bibinfo{author}{\bibfnamefont{J.}~\bibnamefont{Mizrahi}},
  \bibinfo{author}{\bibfnamefont{C.}~\bibnamefont{Senko}}, \bibnamefont{and}
  \bibinfo{author}{\bibfnamefont{C.}~\bibnamefont{Monroe}},
  \bibinfo{journal}{Physical Review Letters} \textbf{\bibinfo{volume}{104}},
  \bibinfo{pages}{140501} (\bibinfo{year}{2010}).

\bibitem[{\citenamefont{Colombe et~al.}(2014)\citenamefont{Colombe, Slichter,
  Wilson, Leibfried, and Wineland}}]{colombe2014single}
\bibinfo{author}{\bibfnamefont{Y.}~\bibnamefont{Colombe}},
  \bibinfo{author}{\bibfnamefont{D.~H.} \bibnamefont{Slichter}},
  \bibinfo{author}{\bibfnamefont{A.~C.} \bibnamefont{Wilson}},
  \bibinfo{author}{\bibfnamefont{D.}~\bibnamefont{Leibfried}},
  \bibnamefont{and} \bibinfo{author}{\bibfnamefont{D.~J.}
  \bibnamefont{Wineland}}, \bibinfo{journal}{Optics express}
  \textbf{\bibinfo{volume}{22}}, \bibinfo{pages}{19783} (\bibinfo{year}{2014}).

\bibitem[{\citenamefont{Mount et~al.}(2016)\citenamefont{Mount, Gaultney,
  Vrijsen, Adams, Baek, Hudek, Isabella, Crain, van Rynbach, Maunz
  et~al.}}]{mount2016scalable}
\bibinfo{author}{\bibfnamefont{E.}~\bibnamefont{Mount}},
  \bibinfo{author}{\bibfnamefont{D.}~\bibnamefont{Gaultney}},
  \bibinfo{author}{\bibfnamefont{G.}~\bibnamefont{Vrijsen}},
  \bibinfo{author}{\bibfnamefont{M.}~\bibnamefont{Adams}},
  \bibinfo{author}{\bibfnamefont{S.-Y.} \bibnamefont{Baek}},
  \bibinfo{author}{\bibfnamefont{K.}~\bibnamefont{Hudek}},
  \bibinfo{author}{\bibfnamefont{L.}~\bibnamefont{Isabella}},
  \bibinfo{author}{\bibfnamefont{S.}~\bibnamefont{Crain}},
  \bibinfo{author}{\bibfnamefont{A.}~\bibnamefont{van Rynbach}},
  \bibinfo{author}{\bibfnamefont{P.}~\bibnamefont{Maunz}},
  \bibnamefont{et~al.}, \bibinfo{journal}{Quantum Information Processing}
  \textbf{\bibinfo{volume}{15}}, \bibinfo{pages}{5281} (\bibinfo{year}{2016}).

\bibitem[{\citenamefont{Leibfried et~al.}(2003)\citenamefont{Leibfried,
  DeMarco, Meyer, Lucas, Barrett, Britton, Itano, Jelenkovi{\'c}, Langer,
  Rosenband et~al.}}]{leibfried2003experimental}
\bibinfo{author}{\bibfnamefont{D.}~\bibnamefont{Leibfried}},
  \bibinfo{author}{\bibfnamefont{B.}~\bibnamefont{DeMarco}},
  \bibinfo{author}{\bibfnamefont{V.}~\bibnamefont{Meyer}},
  \bibinfo{author}{\bibfnamefont{D.}~\bibnamefont{Lucas}},
  \bibinfo{author}{\bibfnamefont{M.}~\bibnamefont{Barrett}},
  \bibinfo{author}{\bibfnamefont{J.}~\bibnamefont{Britton}},
  \bibinfo{author}{\bibfnamefont{W.~M.} \bibnamefont{Itano}},
  \bibinfo{author}{\bibfnamefont{B.}~\bibnamefont{Jelenkovi{\'c}}},
  \bibinfo{author}{\bibfnamefont{C.}~\bibnamefont{Langer}},
  \bibinfo{author}{\bibfnamefont{T.}~\bibnamefont{Rosenband}},
  \bibnamefont{et~al.}, \bibinfo{journal}{Nature}
  \textbf{\bibinfo{volume}{422}}, \bibinfo{pages}{412} (\bibinfo{year}{2003}).

\bibitem[{\citenamefont{Blume-Kohout et~al.}(2017)\citenamefont{Blume-Kohout,
  Gamble, Nielsen, Rudinger, Mizrahi, Fortier, and
  Maunz}}]{blume2017demonstration}
\bibinfo{author}{\bibfnamefont{R.}~\bibnamefont{Blume-Kohout}},
  \bibinfo{author}{\bibfnamefont{J.~K.} \bibnamefont{Gamble}},
  \bibinfo{author}{\bibfnamefont{E.}~\bibnamefont{Nielsen}},
  \bibinfo{author}{\bibfnamefont{K.}~\bibnamefont{Rudinger}},
  \bibinfo{author}{\bibfnamefont{J.}~\bibnamefont{Mizrahi}},
  \bibinfo{author}{\bibfnamefont{K.}~\bibnamefont{Fortier}}, \bibnamefont{and}
  \bibinfo{author}{\bibfnamefont{P.}~\bibnamefont{Maunz}},
  \bibinfo{journal}{Nature communications} \textbf{\bibinfo{volume}{8}},
  \bibinfo{pages}{1} (\bibinfo{year}{2017}).

\bibitem[{\citenamefont{Noek et~al.}(2013)\citenamefont{Noek, Vrijsen,
  Gaultney, Mount, Kim, Maunz, and Kim}}]{noek2013high}
\bibinfo{author}{\bibfnamefont{R.}~\bibnamefont{Noek}},
  \bibinfo{author}{\bibfnamefont{G.}~\bibnamefont{Vrijsen}},
  \bibinfo{author}{\bibfnamefont{D.}~\bibnamefont{Gaultney}},
  \bibinfo{author}{\bibfnamefont{E.}~\bibnamefont{Mount}},
  \bibinfo{author}{\bibfnamefont{T.}~\bibnamefont{Kim}},
  \bibinfo{author}{\bibfnamefont{P.}~\bibnamefont{Maunz}}, \bibnamefont{and}
  \bibinfo{author}{\bibfnamefont{J.}~\bibnamefont{Kim}},
  \bibinfo{journal}{Optics letters} \textbf{\bibinfo{volume}{38}},
  \bibinfo{pages}{4735} (\bibinfo{year}{2013}).

\bibitem[{\citenamefont{Wineland et~al.}(1998)\citenamefont{Wineland, Monroe,
  Itano, Leibfried, King, and Meekhof}}]{wineland1998experimental}
\bibinfo{author}{\bibfnamefont{D.~J.} \bibnamefont{Wineland}},
  \bibinfo{author}{\bibfnamefont{C.}~\bibnamefont{Monroe}},
  \bibinfo{author}{\bibfnamefont{W.~M.} \bibnamefont{Itano}},
  \bibinfo{author}{\bibfnamefont{D.}~\bibnamefont{Leibfried}},
  \bibinfo{author}{\bibfnamefont{B.~E.} \bibnamefont{King}}, \bibnamefont{and}
  \bibinfo{author}{\bibfnamefont{D.~M.} \bibnamefont{Meekhof}},
  \bibinfo{journal}{Journal of Research of the National Institute of Standards
  and Technology} \textbf{\bibinfo{volume}{103}}, \bibinfo{pages}{259}
  (\bibinfo{year}{1998}).

\bibitem[{\citenamefont{Deslauriers et~al.}(2006)\citenamefont{Deslauriers,
  Olmschenk, Stick, Hensinger, Sterk, and Monroe}}]{DeslauriersPRL2006}
\bibinfo{author}{\bibfnamefont{L.}~\bibnamefont{Deslauriers}},
  \bibinfo{author}{\bibfnamefont{S.}~\bibnamefont{Olmschenk}},
  \bibinfo{author}{\bibfnamefont{D.}~\bibnamefont{Stick}},
  \bibinfo{author}{\bibfnamefont{W.~K.} \bibnamefont{Hensinger}},
  \bibinfo{author}{\bibfnamefont{J.}~\bibnamefont{Sterk}}, \bibnamefont{and}
  \bibinfo{author}{\bibfnamefont{C.}~\bibnamefont{Monroe}},
  \bibinfo{journal}{Phys. Rev. Lett.} \textbf{\bibinfo{volume}{97}},
  \bibinfo{pages}{103007} (\bibinfo{year}{2006}).

\bibitem[{\citenamefont{Hite et~al.}(2012)\citenamefont{Hite, Colombe, Wilson,
  Brown, Warring, J\"ordens, Jost, McKay, Pappas, Leibfried
  et~al.}}]{HitePRL2012}
\bibinfo{author}{\bibfnamefont{D.~A.} \bibnamefont{Hite}},
  \bibinfo{author}{\bibfnamefont{Y.}~\bibnamefont{Colombe}},
  \bibinfo{author}{\bibfnamefont{A.~C.} \bibnamefont{Wilson}},
  \bibinfo{author}{\bibfnamefont{K.~R.} \bibnamefont{Brown}},
  \bibinfo{author}{\bibfnamefont{U.}~\bibnamefont{Warring}},
  \bibinfo{author}{\bibfnamefont{R.}~\bibnamefont{J\"ordens}},
  \bibinfo{author}{\bibfnamefont{J.~D.} \bibnamefont{Jost}},
  \bibinfo{author}{\bibfnamefont{K.~S.} \bibnamefont{McKay}},
  \bibinfo{author}{\bibfnamefont{D.~P.} \bibnamefont{Pappas}},
  \bibinfo{author}{\bibfnamefont{D.}~\bibnamefont{Leibfried}},
  \bibnamefont{et~al.}, \bibinfo{journal}{Phys. Rev. Lett.}
  \textbf{\bibinfo{volume}{109}}, \bibinfo{pages}{103001}
  (\bibinfo{year}{2012}).

\bibitem[{\citenamefont{Ozeri et~al.}(2007)\citenamefont{Ozeri, Itano,
  Blakestad, Britton, Chiaverini, Jost, Langer, Leibfried, Reichle, Seidelin
  et~al.}}]{ozeri2007errors}
\bibinfo{author}{\bibfnamefont{R.}~\bibnamefont{Ozeri}},
  \bibinfo{author}{\bibfnamefont{W.~M.} \bibnamefont{Itano}},
  \bibinfo{author}{\bibfnamefont{R.}~\bibnamefont{Blakestad}},
  \bibinfo{author}{\bibfnamefont{J.}~\bibnamefont{Britton}},
  \bibinfo{author}{\bibfnamefont{J.}~\bibnamefont{Chiaverini}},
  \bibinfo{author}{\bibfnamefont{J.~D.} \bibnamefont{Jost}},
  \bibinfo{author}{\bibfnamefont{C.}~\bibnamefont{Langer}},
  \bibinfo{author}{\bibfnamefont{D.}~\bibnamefont{Leibfried}},
  \bibinfo{author}{\bibfnamefont{R.}~\bibnamefont{Reichle}},
  \bibinfo{author}{\bibfnamefont{S.}~\bibnamefont{Seidelin}},
  \bibnamefont{et~al.}, \bibinfo{journal}{Physical Review A}
  \textbf{\bibinfo{volume}{75}}, \bibinfo{pages}{042329}
  (\bibinfo{year}{2007}).

\bibitem[{sup()}]{supplemental1}
\bibinfo{note}{See Supplemental Material [url] for details of projection
  optics, the MS gate simulation and the coherence time measurements, which
  includes Refs. [45-47].}

\bibitem[{\citenamefont{Hayes}(2012)}]{hayes2012remote}
\bibinfo{author}{\bibfnamefont{D.~L.} \bibnamefont{Hayes}}, Ph.D. thesis
  (\bibinfo{year}{2012}).

\bibitem[{\citenamefont{Lindblad}(1976)}]{lindblad1976generators}
\bibinfo{author}{\bibfnamefont{G.}~\bibnamefont{Lindblad}},
  \bibinfo{journal}{Communications in Mathematical Physics}
  \textbf{\bibinfo{volume}{48}}, \bibinfo{pages}{119} (\bibinfo{year}{1976}).

\bibitem[{\citenamefont{Gardiner et~al.}(2004)\citenamefont{Gardiner, Zoller,
  and Zoller}}]{gardiner2004quantum}
\bibinfo{author}{\bibfnamefont{C.}~\bibnamefont{Gardiner}},
  \bibinfo{author}{\bibfnamefont{P.}~\bibnamefont{Zoller}}, \bibnamefont{and}
  \bibinfo{author}{\bibfnamefont{P.}~\bibnamefont{Zoller}},
  \emph{\bibinfo{title}{Quantum noise: a handbook of Markovian and
  non-Markovian quantum stochastic methods with applications to quantum
  optics}} (\bibinfo{publisher}{Springer Science \& Business Media},
  \bibinfo{year}{2004}).

\end{thebibliography}

\onecolumngrid
\newpage
\twocolumngrid

\section{APPENDIX}
\section{\label{sec:simulation} APPENDIX A: FM MS gate simulation}

The study of imperfections and error mechanisms in a FM MS gate provides insights into whether the limitations are of fundamental nature, or technical challenges. We use a combination of Monte Carlo simulation, Schr\"{o}dinger equation simulation, and master equation simulation for the study of these error mechanisms. We use the Schr\"{o}dinger equation to study coherent systematic errors without consideration of dissipation, e.g., detuning errors, drifts of laser intensity and calibration drifts. The Hamiltonian of the MS evolution of the $j$th motional mode with no modulation is written as \cite{sorensen1999quantum,sorensen2000entanglement,hayes2012remote}
\begin{widetext}
\begin{eqnarray}
\label{eq1}\hat{H}(t)_{j,MS} = \frac{i}{2} \sum_{n=1,2} \eta_{j}^{(n)}\hat{\sigma}_{+}^{(n)}\left(\Omega_{r}^{(n)} \hat{a}_j e^{i \phi_{r}-i \delta_{j,r}^{(n)}t}+\Omega_{b}^{(n)}\hat{a}_j^{\dag}e^{i \phi_{b}-i \delta_{j,b}^{(n)}t} \right) + h.c.
\end{eqnarray}
\end{widetext}
where $\Omega_{r}^{(1)}$, $\Omega_{b}^{(1)}$, $\Omega_{r}^{(2)}$ and $\Omega_{b}^{(2)}$ are the Rabi frequencies of red and blue sideband transitions for the two target ions, $\delta_{j,r}^{(1)}$, $\delta_{j,b}^{(1)}$, $\delta_{j,r}^{(2)}$, and $\delta_{j,b}^{(2)}$ are the detunings for the $j$th motional mode, $\phi_r$ and $\phi_b$ are the laser phases of the red and blue tone, respectively. With the expansion in Eq. (\ref{eq1}), we can simulate number of error mechanisms: power imbalance on two target ions, power imbalance on red and blue tones, and detuning imbalance due to Stark shift. For the full MS evolution, the modes are sequentially simulated to minimize the computing resource. We only save the spin state result for the next round of simulation. The Hamiltonian of different modes commute when $\Omega_{r}^{(1)} = \Omega_{b}^{(1)}$ and $\Omega_{r}^{(2)} = \Omega_{b}^{(2)}$, which is a reasonable assumption in the MS gate. For the evolution of discrete segments in FM gates, we sequentially simulate every segment to obtain the final state.

We use a master equation \cite{gardiner2004quantum} to simulate an open-quantum system considering multiple dissipative error mechanisms: motional heating, motional dephasing, and laser dephasing. The master equation is written in Lindblad form~\cite{lindblad1976generators}
\begin{equation*}
    \frac{d\hat{\rho}}{d t} = \frac{1}{i \hbar}[\hat{H},\hat{\rho}]+\sum_j\left( \hat{L}_j\hat{\rho} \hat{L}_j^{\dag} -\frac{1}{2} \hat{L}_j^{\dag} \hat{L}_j \hat{\rho} -\frac{1}{2} \hat{\rho} \hat{L}_j^{\dag} \hat{L}_j \right),
\end{equation*}
where $\rho$ is the density matrix of the system, $H$ is the Hamiltonian of the MS gate, $\hat{L}_j$ is the Lindblad operator for the $j$th decoherence process. The motional dephasing can be described by the Lindblad operator of the form $\hat{L}_m = \sqrt{\frac{2}{\tau_m}}\hat{a}^{\dag}\hat{a}$, where $\tau_m$ is the motional coherence time. The anomalous heating can be described by $\hat{L}_{+} = \sqrt{\Gamma}\hat{a}^{\dag}$ and $\hat{L}_{-} = \sqrt{\Gamma}\hat{a}$, where $\Gamma$ is the heating rate. For these two operators, we sequentially simulate the evolution of each mode, then combine them to obtain the final state. The master equation simulations represent the full density matrix representation for a truncated state space of two qubits and one motional mode truncated to the first 13 Fock states ($n \leq 12$). The laser dephasing can be described by the Lindblad operator of $\hat{L}_l = \sqrt{1/\tau_l}(\hat{\sigma}_{z}^{(1)}+\hat{\sigma}_{z}^{(2)})$, where $\tau_l$ is the laser coherence time. For this Lindblad operator, we perform a full master equation simulation with all motional modes and spin states included. We truncate the far off-resonance motional modes, which have a smaller motional excitation, to smaller Fock states to save on computational resources. For the stochastic noise, we also combined the simulation with Monte Carlo method. The simulations are performed using \textit{Mathematica} software.

\section{\label{sec:simulation} APPENDIX B: Laser and motional coherence time}

\begin{figure}[!htb]
\center{\includegraphics[width=0.48 \textwidth]{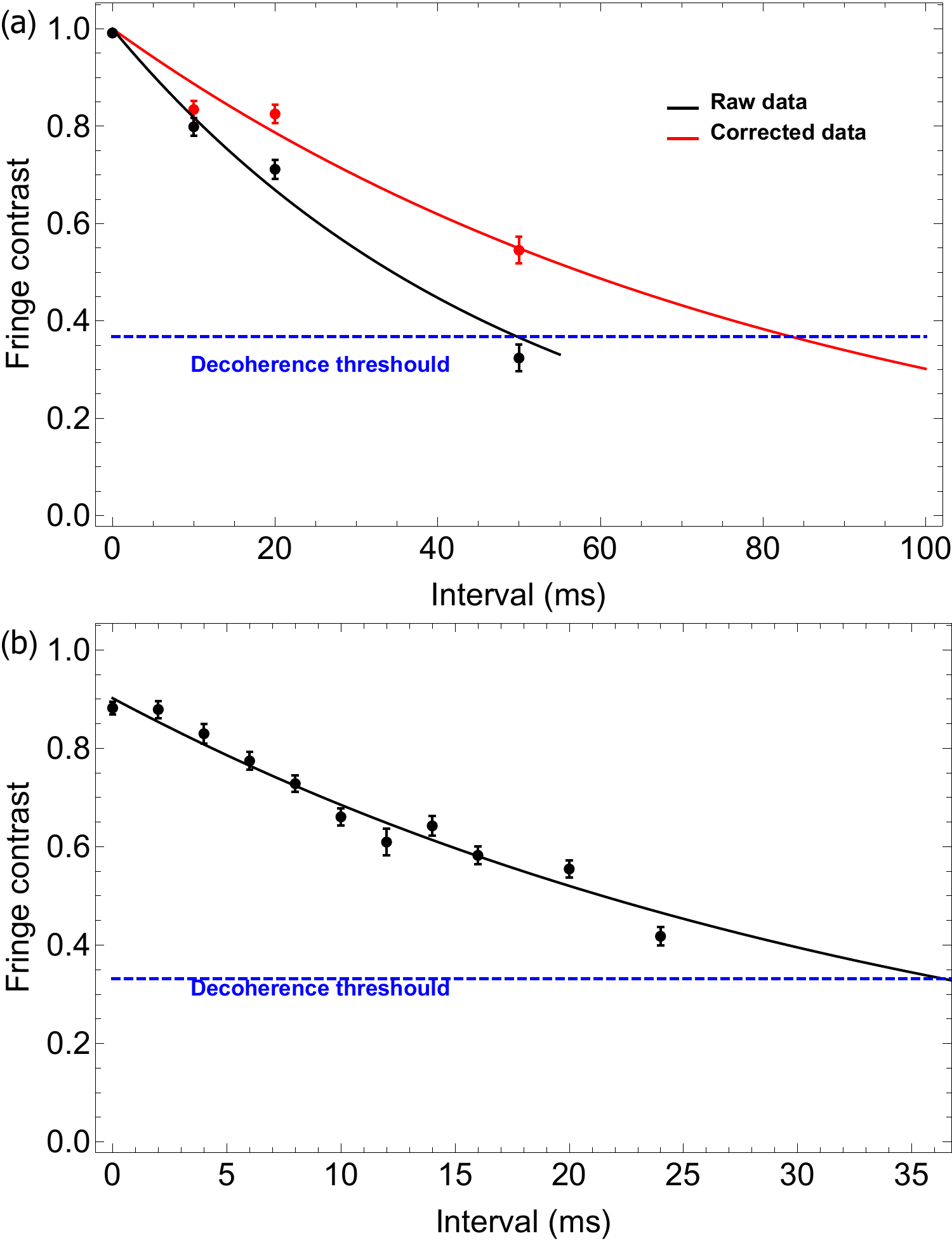}}
\caption{\label{fig:laserCoherence} \textbf{(a)} The black and red dots are the data of Ramsey contrasts depending on interval time without and with correction for Debye-Waller effect, respectively. The fitted coherence times are $49.7 \pm 45$ ms and $83.3\pm 11.5$ ms, respectively. Error bars are standard deviations. \textbf{(b)} The Ramsey contrast data for the motional coherence measurement. The fitted coherence time is $36.3 \pm 2.3$ ms. Error bars are standard deviations.}
\end{figure}

We perform Ramsey interferometry measurements to obtain the laser and motional coherence time, as described in the main text. The Rabi frequencies of the two transitions we used are affected by the Debye-Waller effect \cite{wineland1998experimental}. The effect on the carrier transition is a reduction in the Rabi frequency:
\begin{align}
    \Omega' &= \Omega \prod_{j = 1,2}e^{-1/2\eta_j^2} \mathcal{L}_{n_j}(\eta_j^2)\\
    &\approx \Omega \prod_{j = 1,2}\mathcal{L}_{n_j}(\eta_j^2),
\end{align}
where $\Omega$ is the original Rabi frequency of the carrier transition, $\eta_j$ is the Lamb Dicke parameter of two radial modes, $n_j$ is the vibrational number of radial modes, $\mathcal{L}_{n_j}$ is the $n_j$th Laguerre polynomial. The principal axes are rotated to about $45^\circ$ to the trap surface, which lead to negligible coupling to one of the radial modes and $\eta = 0.1$ for the other radial mode. As a result, we only need to consider the Debye-Waller effect of one mode. If we assume the state of the radial motional mode is thermally distributed with average phonon number $\overline{n}$:
\begin{align}
    P_{n} &= \frac{\overline{n}^{n}}{(1+\bar{n})^{n+1}},
\end{align}
the mean Rabi frequency is given by
\begin{align}
    \label{eq:DebyeWaller}
    \overline{\Omega'} \approx \Omega \sum_{n=0}^\infty P_{n}\mathcal{L}_{n}(\eta^2).
\end{align}
The second $\pi/2$ rotation in the Ramsey measurement with an average phonon number $\overline{n_j}$ will cause a contrast decay of $\sin (\pi\overline{\Omega'}/(2\Omega))^2$. We calculate $\overline{n}$ using the interval time and a heating rate of $1000$ phonon/s, which was the case for this single ion dataset. We then calculated the corrected contrast base on Eq. (\ref{eq:DebyeWaller}). As shown in Fig. \ref{fig:laserCoherence}(a), the black and red dots are the measured Ramsey contrasts depending on interval time without and with the correction, respectively. The black and red lines are exponential fits to the data from which the laser coherence time is extracted.

The Ramsey interferomety measurement for the motional coherence time is taken with the zig-zag mode of a 7-ion chain, which has negligible heating rate. The data and exponential fit is shown in Fig. \ref{fig:laserCoherence} (b). The imperfect contrast at the beginning is due to a small drift of the mode frequency. The exponential fit yields the motional coherence time. 

\section{\label{sec:optics} APPENDIX C: Projection and Beam Combining Optics}
A lens focuses the two parallel individual beams onto the first set of MEMS mirrors and, in combination with a concave mirror, projects the Gaussian beam waist onto the second set of MEMS mirrors. The concave mirror also functions as a 2\textit{f}-2\textit{f} imaging system to image the beams reflecting off the first MEMS mirrors onto the second MEMS mirrors. The mirror tilting angle can be precisely controlled by an actuating voltage with a switching time of \SI{\sim 5}{\micro s}. A lens placed a focal length \textit{f} away from the second MEMS mirrors converts the tilt of the beams reflecting off the mirrors to a parallel beam shift with the beam waist situated in the Fourier plane. The Fourier plane is then demagnified and projected before the high NA lens which finally images the beam waist onto the qubits. The projection optics also overlaps the two individual beam paths.

\end{document}